\documentclass[epj,referee]{svjour}
\usepackage{graphics}
\begin{document}
\title{Loss induced collective subradiant Dicke behaviour in a multiatom sample}

\author{S. Nicolosi, A. Napoli, A. Messina}
\offprints{}          
\institute{INFM, MIUR and Dipartimento di Scienze Fisiche ed
Astronomiche, Universit\`{a} di Palermo, via Archirafi 36, 90123
Palermo, Italy, Tel/Fax: +39 091 6234243; E-mail:
messina@fisica.unipa.it}
\date{Received: date / Revised version: date}
%
\abstract{ The exact dynamics of $N$ two-level atoms coupled to a
common electromagnetic bath and closely located inside a lossy
cavity is reported. Stationary radiation trapping effects are
found and very transparently interpreted in the context of our
approach. We prove that initially injecting one excitation only in
the $N$ atoms-cavity system, loss mechanisms asymptotically drive
the matter sample toward a long-lived collective subradiant Dicke
state. The role played by the closeness of the $N$ atoms with
respect to such a cooperative behavior is brought to light and
carefully discussed.
\PACS{
      {03.65.Yz}{Decoherence, open systems, quantum statistical methods}   \and
      {03.67.Mn}{Entanglement production and manipulation}\and
      {42.50.Fx}{Cooperative phenomena in quantum optical systems}
     } 
} 
\maketitle
\section{Introduction}
\label{intro} It is well known that entangled states of two or
more particles give rise to quantum phenomena that cannot be
explained in classical terms. The concept of entanglement was
indeed early recognized as the characteristic trait of the quantum
theory itself. For this reason much interest has been devoted by
many physicists, both theoreticians and experimentalists, toward
the possibility of generating entangled states of bipartite or
multipartite systems. To produce and to be able to modify at will
the degree of entanglement stored in a system is indeed a
desirable target to better capture fundamental aspects of the
quantum world. Over the last decade, moreover, it has been
recognized that the peculiar properties of the entangled states,
both pure and mixed, can be usefully exploited as an effective
resource in the context of quantum information and computation
processing \cite{Beige03,BeigePRA03,Cinesi,Cinesi2}. Such a
research realm has attracted much interest since it becomes clear
that quantum computers are, at least in principle, able to solve
very hard computational problems more efficiently than classical
logic-based ones. The realization of quantum computation protocols
suffers anyway the difficulty of isolating a quantum mechanical
system from its environment. Very recently, however, nearly
decoherence-free quantum gates have been proposed by exploiting,
rather than countering, the same dissipation mechanisms
\cite{Pellizzari,Cirac,Shneider,Yang,Hagley,Foldi,Beige99,Beige00}.
The main requirement to achieve this goal is the existence of a
decoherence-free subspace for the system under consideration. In
the same spirit it has been recognized that transient entanglement
between distant atoms can be induced by atomic spontaneous decay
\cite{Ficek03} or by cavity losses \cite{Jakobczyk}. In ref.
\cite{Jakobczyk} it has also been demonstrated that asymptotic
entangled states of two closely separated two-level atoms in free
space can be created as conseguence of the spontaneous emission
process.

In this paper we present a new path along which loss mechanisms
act {\sl constructively} inducing a collective Dicke behavior in a
multiatom sample. In refs. \cite{Yu,Hong} multistep schemes to
generate a set of Dicke states of multi $\Lambda$-type three level
atoms are reported. In these procedures the key point is the
possibility of successfully incorporating the presence of cavity
losses in the theory, neglecting on the contrary atomic
spontaneous emission.

In order to reach our scope we consider a material system of $N$
identical two-level atoms closely placed inside a resonant bad
cavity taking also into account, from the very beginning, the
coupling between each atom and the external world. Exactly solving
the master equation governing the dynamics of the system under
scrutiny, supposing that only one excitation has been initially
injected in it, we show that the system of the $N$ two-level atoms
may be guided, with appreciable probability, toward a nontrivial
stationary condition described as a Dicke state having the form
$\mid S,-S\rangle $ with $S=\frac{N-2}{2}$, $\mathbf{S}$ being the
total Pauli spin operator of the atomic sample. In addition,
exploiting the knowledge of the exact temporal evolution of the
matter-cavity reduced density matrix, we propose an analytical
route to follow up some physically transparent aspects
characterizing the entanglement building up process in the passage
from a chosen totally uncorrelated  initial  situation to the
manifestly entangled asymptotic one. The treatment followed in our
paper enables to catch the physical origin of the stationary
collective Dicke behavior of the system. In addition it has the
virtue to provide a transparent way to understand the key role
played by the loss mechanisms and by the closeness of the atoms in
the phenomena brought to light. The paper is organized as follows.
The next section is devoted to an accurate presentation of our
physical model and to the formulation of the relative master
equation for the matter-cavity reduced density operator. An
appropriate unitary change of this operator variable provides, in
section 3, the mathematical key tool for exactly solving a Cauchy
problem in the one-excitation subspace of $N$ atoms-resonator
Hilbert space. The two successive sections contain the main
results of this paper. The entanglement formation process is
addressed in section 4 studying the time evolution of the Wootters
concurrence \cite{Wootters97,Wootters98} relative to a generic
pair of two-level atoms. Section 5, in turn, brings to light the
occurrence of stationary collective Dicke subradiant behaviour of
our matter subsystem. The last section contains some final remarks
as well as a discussion on the experimental implementation of the
physical conditions assumed in the paper.

\section{The physical system and its master equation}
\label{sec:1} As previously said, our system consists of $N$
identical two-level atoms closely located within a single-mode
cavity. Indicate the atomic frequency transition and the cavity
mode frequency by $\omega_0$ and $\omega$ respectively and suppose
$\omega_0\sim\omega$. Assume, in addition, that all the conditions
under which the interaction between each atom and the cavity field
is well described by a Jaynes Cummings (JC) model, are satisfied
\cite{Jaynes}. Thus, the unitary time evolution of the system we
are considering is governed by the following hamiltonian:
\begin{equation}\label{HAC}
  H_{AC}=\hbar \omega
  \alpha^{\dag}\alpha+\hbar\frac{\omega_0}{2}\sum_{i=1}^N\sigma_z^{(i)}+
  \hbar\sum_{i=1}^N[\varepsilon^{(i)} \alpha\sigma_+^{(i)}+h.c.]
\end{equation}
In eq.\ (\ref{HAC}) $\alpha$ and $\alpha^{\dag}$ denote the
single-mode cavity field annihilation and creation operators
respectively, whereas $\sigma_z^{(i)}$, $\sigma_{\pm}^{(i)}$
$(i=1,...N)$ are the Pauli operators of the $i$-th atom. The
coupling constant between the $i-th$ atom and the cavity is
denoted by $\varepsilon^{(i)}$.

It is easy to demonstrate that the excitation number operator
$\hat{N}$ defined as
$\hat{N}=\alpha^{\dag}\alpha+\frac{1}{2}\sum_{i=1}^N\sigma_z^{(i)}
+\frac{N}{2}$ is a constant of motion being $[\hat{N},H_{AC}]=0$.
Thus, preparing the physical system at $t=0$ in a state with a
well defined number of excitations $N_e$, its dynamics is confined
in a finite-dimensional Hilbert subspace singled out by
 this  eigenvalue of $\hat{N}$. In a realistic
situation, however, the system we are considering is subjected to
two important sources of decoherence. The first one is undoubtedly
related to the fact that photons can leak out through the cavity
mirrors due to the coupling of the resonator mode to the free
radiation field outside the cavity. Moreover the atoms present
inside the resonator can spontaneously emit photons into
non-cavity field modes. The microscopic hamiltonian  taking into
account these loss mechanisms may be written in the form
\cite{Haroche}
\begin{equation}\label{Htot}
  H=H_{AC}+H_R+H_{AR}+H_{CR}
\end{equation}
where
\begin{equation}\label{HR}
 H_R=\hbar\sum_{\mathbf{k},\lambda}\omega_{\mathbf{k}, \lambda}[c_{\mathbf{k}, \lambda}^{\dag}c_{\mathbf{k}, \lambda}
 + \tilde{c}_{\mathbf{k}, \lambda}^{\dag} \tilde{c}_{\mathbf{k}, \lambda}]
\end{equation}
is the hamiltonian relative to the environment,
\begin{equation}\label{HAR}
  H_{AR}=\sum_{\mathbf{k},\lambda,i}[g_{\mathbf{k},\lambda}^{(i)}\tilde{c}_{\mathbf{k},\lambda}\sigma_+^{(i)}+h.c.]
\end{equation}
describes the interaction between the atomic sample and the bath
and, finally,
\begin{equation}\label{HCR}
  H_{CR}=\sum_{\mathbf{k},\lambda}[s_{\mathbf{k},\lambda}c_{\mathbf{k},\lambda}\alpha^{\dag}+h.c.]
\end{equation}
represents the coupling between the environment and the cavity
field. In eq.\ (\ref{HR}) we have assumed, as usual
\cite{Beige00}, that the two subsystems, the $N$ atoms and the
single-mode cavity, see two different reservoirs. In eq.\
(\ref{HR})-\ (\ref{HCR}) the boson operators relative to the
atomic bath are denoted by $\{ \tilde{c}_{\mathbf{k},\lambda},
\tilde{c}_{\mathbf{k},\lambda}^{\dag}\}$ whereas
$c_{\mathbf{k},\lambda}, c_{\mathbf{k},\lambda}^{\dag}$ are the
$(\mathbf{k},\lambda)$ mode annihilation and creation operators
respectively of the cavity environment. Moreover, the coupling
constants $\{s_{\mathbf{k},\lambda}\}$ are phenomenological
parameters whereas
\begin{equation}\label{gklambda}
  g_{\mathbf{k},\lambda}^{(i)}=-i(\frac{2\pi\hbar\omega_0^2}{V\omega_\mathbf{k}})
  ^\frac{1}{2}(\mathbf{e}_{\mathbf{k}\lambda} \cdot \mathbf{d} )e^{i\mathbf{k}\cdot\mathbf{r}_i}
\end{equation}
stems from a dipole atom-field coupling \cite{Leonardi}. In eq.\
(\ref{gklambda}) $\mathbf{e}_{\mathbf{k}\lambda}$ represents the
polarization vector of the atomic thermal bath $(\mathbf{k}
\lambda)$ mode of frequency $\omega_\mathbf{k}$,
 $V$ its effective volume, $\mathbf{d}$ is the electric dipole
 matrix element between the two atomic levels and, finally,
 $\mathbf{r}_i$ is the position of the $i$-th atom. Indicate now by $\mathbf{\hat{d}}$
 and $\mathbf{\hat{r}}_{ij}$ two unit vectors
along the atomic transition dipole moment and the atomic distance
$\mathbf{r}_{ij}=\mathbf{r}_i-\mathbf{r}_j$, respectively. Our
microscopic hamiltonian model\ (\ref{Htot}) does not take into
account dipole-dipole static interactions among the $N$ atoms
making in this way $H$ invariant under the exchange of two
arbitrary atoms when $r_{ij}<<\frac{c}{\omega_0}$ for any $i$ and
$j$. The two hypotheses implying such a permutational symmetry
property are, in general, contradictory
\cite{Friedberg,Crubellier}. However, notwithstanding the
conceptual difficulties connected with these assumptions, they are
commonly adopted \cite{Ficek03,Jakobczyk,Carmichael,Ficek02} for
the sake of simplicity.

Following standard procedures based on the Rotating-Wave and
Born-Markov approximations \cite{F.Petruccione,W.H.Louisell}, it
is possible to prove that the reduced density operator $\rho_{AC}$
relative to the bipartite system composed by the $N$ atoms
subsystem and the single-mode cavity, evolves  nonunitarily in
time in accordance with the following Lindbland master equation
\cite{Agarval}:
\begin{equation}\label{m-e}
  \dot{\rho}_{AC}=-\frac{i}{\hbar} [H_{AC}+H_{LS},\rho_{AC}]+
  \mathcal{L}_f\rho_{AC}+ \mathcal{L}_A\rho_{AC}
\end{equation}
where
\begin{equation}
 H_{LS}=\sum_{i,j}\Omega_{ij}\sigma_i^+ \sigma_j^-
\end{equation}
\begin{eqnarray}\label{Omega}
  \Omega_{ij}&=&\frac{3}{4}\Gamma\{[(\mathbf{\hat{d}}\cdot\mathbf{\hat{r}_{ij}})^2-1]
  c\frac{\cos(\frac{\omega_0}{c}r_{ij})}{\omega_0r_{ij}}+ \\
  \nonumber
  &+&[1-3(\mathbf{\hat{d}}\cdot\mathbf{\hat{r}_{ij}})^2][c^2\frac{\sin
  (\frac{\omega_0}{c}r_{ij})}
  {(\omega_0r_{ij})^2}+c^3\frac{\cos
  (\frac{\omega_0}{c}r_{ij})}
  {(\omega_0r_{ij})^3}] \}
\end{eqnarray}
\begin{equation}\label{Lf}
 \mathcal{L}_f\rho_{AC}=
k(2\alpha\rho_{AC}\alpha^{\dag}-\alpha^{\dag}\alpha\rho_{AC}-\rho_{AC}\alpha^{\dag}\alpha)
\end{equation}
\begin{eqnarray}\label{La}
 && \mathcal{L}_A\rho_{AC}= \\ \nonumber
  &&\sum_{i=1}^{N}{\Gamma_{ii}(2\sigma_{-}^{(i)}\rho_{AC}\sigma_{+}^{(i)}-
  \sigma_{+}^{(i)}\sigma_{-}^{(i)}\rho_{AC}-\rho_{AC}\sigma_{+}^{(i)}\sigma_{-}^{(i)})}+\\
  \nonumber
  &&\sum_{i,j=1(i\neq j)}^{N}{\Gamma_{i,j}(2\sigma_{-}^{(i)}\rho_{AC}\sigma_{+}^{(j)}-
  \sigma_{+}^{(i)}\sigma_{-}^{(j)}\rho_{AC}-\rho_{AC}\sigma_{+}^{(i)}\sigma_{-}^{(j)})}
\end{eqnarray}

We point out that both the two baths entering in our model are
supposed in thermal states  at $T=0$ and that, when
$r_{ij}<<\frac{c}{\omega_0}$, $\Omega_{ij}$ tends toward a
limiting value $\Omega_L$ independent from $i$ and $j$
\cite{Agarval}.

The decay rate $k$ appearing in eq.\ (\ref{Lf}) is given by
\begin{equation}\label{k}
  k=\sum_{\mathbf{k}\lambda}{\vert s_{\mathbf{k}\lambda}\vert ^2
  \delta(\omega_{\mathbf{k}}-\omega)}
\end{equation}
Moreover in eq.\ (\ref{La}) the $N^2$ coupling constants
\begin{equation}\label{gamma}
  \Gamma_{ii}\equiv\Gamma=\frac{4\pi \hbar \omega_0^3|\mathbf{d}|^2}{3  c^3}
\end{equation}
\begin{equation}\label{gammaij}
  \Gamma_{ij}=\Gamma_{ji}=\Gamma f_{ij} \;\;\;\;i\neq j
\end{equation}
related to the spontaneous emission loss channel, define the
spectral correlation tensor $\mathbf{\Gamma}$
\cite{F.Petruccione}.

In eq.\ (\ref{gammaij}) the function $f_{ij}$ is defined as
follows:
\begin{eqnarray}\label{f}
  f_{ij}&=&\frac{3}{2}\{[1-(\mathbf{\hat{d}}\cdot\mathbf{\hat{r}_{ij}})^2]
  c\frac{\sin(\frac{\omega_0}{c}r_{ij})}{\omega_0r_{ij}}+ \\
  \nonumber
  &+&[1-3(\mathbf{\hat{d}}\cdot \mathbf{\hat{r}_{ij}})^2][c^2\frac{\cos
  (\frac{\omega_0}{c}r_{ij})}
  {(\omega_0r_{ij})^2}-c^3\frac{\sin
  (\frac{\omega_0}{c}r_{ij})}
  {(\omega_0r_{ij})^3}] \}
\end{eqnarray}
It is important to underline that the last term appearing in the
right hand side of equation\ (\ref{La}), is a direct consequence
of the fact that we have considered, from the very beginning, a
$common$ bath for the $N$ atoms. As we shall see, this term is
responsible for cooperative effects among the $N$ atoms leading to
the possibility of generating asymptotic entangled states of the
atomic sample, immune from decoherence. We wish to stress that it
is the nearness of the atoms inside the cavity that imposes the
consideration of a common bath. If, on the other hand, the
distance among the atoms became large enough $(r_{ij}\gg
\frac{c}{\omega_0})$, these cooperative effects, as deducible from
eq.\ (\ref{f}), would disappear so that the dynamics of the system
could be equivalently obtained considering $N$ different
reservoirs, one for each two-level atom. In such a situation the
system would evolve toward its vacuum state with no excitation.

\section{One-excitation exact dynamics}
In what follows we assume that the atoms within the cavity are
located at a distance smaller than the wavelength of the cavity
mode, thus legitimating the henceforth done position
$\varepsilon^{(i)}\equiv\varepsilon$ and $g_{k,\lambda}^{(i)}
\equiv g_{k,\lambda}$ for any $i$. Under this hypothesis we solve
eq.\ (\ref{m-e}) exploiting the unitary operator $U$
\cite{Benivegna88} defined as
\begin{equation}\label{U}
  U=\prod_{i=2}^N U_i
\end{equation}
where
\begin{equation}\label{Ui}
  U_i=e^{\delta_i(\sigma_+^1\sigma_-^i-\sigma_-^1\sigma_+^i)}\;\;\;\;\;i=2,...,N
\end{equation}
with $\delta_i=-\arctan(\frac{1}{\sqrt{i-1}})$ and
$[U,\hat{N}]=0$. It is easy to demonstrate that, if no more than
one excitation is initially stored in the atom-cavity physical
system
\begin{equation}\label{sigmatransf}
U^{\dag}\sigma_{+}^{(i)}
U=\sqrt{\frac{i-1}{i}}\sigma_+^{(i)}-\sum_{l=i+1}^N
\sqrt{\frac{1}{l(l-1)}}\sigma_+^{(l)}+\frac{1}{\sqrt{N}}\sigma_+^{(1)}.
\end{equation}
for $i=1,..,N$ so that
\begin{eqnarray}\label{HACtras}
  \tilde{H}_{AC}\equiv && U^{\dag}H_{AC}U=\\ \nonumber
  &&\hbar \omega
  \alpha^{\dag}\alpha+\hbar\frac{\omega_0}{2}\sum_{i=1}^N\sigma_z^{(i)}+
  \hbar \varepsilon_{eff}[\alpha\sigma_+^{(1)}+h.c.]
\end{eqnarray}
with $\varepsilon_{eff}=\sqrt{N}\varepsilon$. Transforming in the
$N_e=0,1$ excitation subspace the operator variable $\rho_{AC}$
into the new one $\tilde{\rho}_{AC}\equiv U^{\dag}\rho_{AC}U$ and
taking into account that
\begin{equation}\label{Hlstransf}
\tilde{H}_{LS}\equiv U^{\dag}H_{LS}U=\Omega_L
N\sigma_+^{(1)}\sigma_-^{(1)},
\end{equation}
$\Omega_L$ being the common limiting value of $\Omega_{ij}$ when
$r_{ij}$ tends to zero, it is not difficult to convince oneself
that
\begin{eqnarray}\label{m-e-tr}
  \dot{\tilde{\rho}}_{AC}&=&-\frac{i}{\hbar}
  [\tilde{H}_{AC}+\tilde{H}_{LS},\tilde{\rho}_{AC}]\\ \nonumber
  &+&k(2 \alpha\tilde{\rho}_{AC}\alpha^{\dag}-
  \alpha^{\dag}\alpha\tilde{\rho}_{AC}-\tilde{\rho}_{AC}\alpha^{\dag}\alpha)\\
  \nonumber
  &+& N \Gamma(2 \sigma_-^{(1)}\tilde{\rho}_{AC}\sigma_+^{(1)}-
  \sigma_+^{(1)}\sigma_-^{(1)}\tilde{\rho}_{AC}-\tilde{\rho}_{AC}\sigma_+^{(1)}\sigma_-^{(1)})
\end{eqnarray}
in view of eq.\ (\ref{m-e}),\ (\ref{U})-\ (\ref{Hlstransf}).
Comparing eq.\ (\ref{m-e-tr}) with eq.\ (\ref{m-e}) shows that in
the new representation the correspondent spectral correlation
tensor $\tilde{\mathbf{\Gamma}}$ is in diagonal form, moreover
being $\tilde{\Gamma}_{ii}=\Gamma \delta_{1,i}$. The physical
meaning of this peculiar property is that the atomic subsystem in
the transformed representation looses its energy only through the
interaction of the first atom with both the cavity mode and the
environment. Such a behaviour stems from the fact that, in view of
eq.\ (\ref{HACtras}), the other $N-1$ atoms freely evolve being
decoupled either from the cavity field and from the
electromagnetic modes of the thermal bath. It is of relevance to
underline that the form assumed by the terms associated to the
nonunitary evolution, appearing in eq.\ (\ref{m-e-tr}), directly
reflects the main role played by the closeness of the atoms in our
model. It is indeed just this feature which leads, in the
transformed representation, to the existence of $N-1$ collective
atoms immune from spontaneous emission losses and, at the same
time, decoupled from the cavity mode. Thus, to locate the atomic
sample within a linear dimension much shorter than the wavelength
of the cavity mode, introduces an essential permutational atomic
symmetry which is at the origin of a collective replay of the $N$
atoms such that, even in presence of both the proposed dissipation
routes, the matter subsystem may stationarily trap the initial
energy.

Bearing in mind that $[H_{AC},\hat{N}]=0$ and $[U,\hat{N}]=0$, it
is immediate to convince oneself that, if only one excitation
 is initially injected into the
atomic subsystem, whereas the cavity is prepared in its vacuum
state, at a generic time instant $t$, the density operator
$\tilde{\rho}_{AC}$, can have not vanishing matrix elements only
in the Hilbert subspace generated by the following ordered set of
$N+2$ state vectors:
\begin{eqnarray}\label{basis}
  \nonumber && \vert 0 \rangle \vert - \rangle_1 \vert - \rangle_2...\vert -
  \rangle_N \equiv \vert \beta_1 \rangle \\ \nonumber
  &&\vert 0 \rangle \vert - \rangle_1...\vert + \rangle_h...\vert
  -\rangle_N\equiv \vert \beta_{h+1} \rangle \;\;\;\; h=1,..,N\\ \nonumber
  &&\vert 1 \rangle\vert - \rangle_1...\vert - \rangle_N
  \equiv \vert \beta_{N+2} \rangle \\ \nonumber
\end{eqnarray}
where $\vert p \rangle$ $(p=0,1)$ is a number state of the cavity
mode and $\vert + \rangle_h$ $(\vert - \rangle_h)$ denotes the
excited (ground) state of the $h$-th collective atom
$(h=1,...,N)$. Eq.\ (\ref{m-e-tr}) can thus be easily converted
into a system of coupled differential equations involving the
density matrix elements $\tilde{\rho}_{hj}\equiv \langle
\beta_{h}\vert \tilde{\rho}\vert\beta_{j}\rangle$ with
$h,j=1,...N+2$. At this point let's observe that from an
experimental point of view it seems reasonable to think that the
excitation given at $t=0$ to the matter sample can be captured by
$i$-th the atom or by $j$-th with the same probability. In other
words our initial condition must be reasonably represented as
statistical mixture of states $\vert \beta_{h} \rangle$, with
$h\geq 2$, of the form
\begin{eqnarray}\label{ro0}
 \rho_{AC}(0)=\frac{1}{N}\sum_{h=2}^{N+1}\vert \beta_{h}
 \rangle\langle \beta_{h}\vert.
\end{eqnarray}
Exploiting eq.\ (\ref{sigmatransf}) it is possible to verify that
\begin{equation}\label{rotilde0}
\tilde{\rho}_{AC}(0)\equiv \rho_{AC}(0)
\end{equation}
After lengthly and tedious calculations and taking into account
eqs.\ (\ref{ro0})-\ (\ref{rotilde0}), we have exactly determined
the time evolution of each $\tilde{\rho}_{h,j}(t)$ $h,j=1,..,N+2$
finding:
\begin{eqnarray}\label{matrix1}
 \nonumber&&\tilde{\rho}_{AC}(t)=\\&&\left(
\begin{array}{c c c c c c}

 {\scriptstyle\tilde{\rho}_{11}(t)} & {\scriptstyle 0} & {\scriptstyle 0} & ... & {\scriptstyle 0} & {\scriptstyle 0} \\

{\scriptstyle 0} & {\scriptstyle\tilde{\rho}_{22}(t)} & {\scriptstyle 0} & ... & {\scriptstyle 0} & {\scriptstyle\tilde{\rho}_{2,N+2}(t) }\\

{\scriptstyle 0} & {\scriptstyle 0} & {\scriptstyle\tilde{\rho}_{33}(t)} & ... & {\scriptstyle 0} & {\scriptstyle 0} \\

\vdots & \vdots & \vdots & \vdots\vdots\vdots & \vdots & \vdots \\

{\scriptstyle 0} & {\scriptstyle 0}v & . & ... & {\scriptstyle\tilde{\rho}_{N+1,N+1}(t) }& {\scriptstyle 0} \\

{\scriptstyle 0} & {\scriptstyle\tilde{\rho}_{2,N+2}^*(t) }& . &
... & {\scriptstyle 0} & {\scriptstyle\tilde{\rho}_{N+2,N+2}(t)}

\end{array}
\right)
\end{eqnarray}
where $\tilde{\rho}_{i,j}(t)=\tilde{\rho}_{i,j}(0)$ $(3\leq
i,j\leq N+1)$,
$\tilde{\rho}_{1,1}(t)=1-\sum_{i=2}^{N+2}\tilde{\rho}_{i,i}(t)$
and
\begin{eqnarray}\label{rho22}
 \nonumber \tilde{\rho}_{22}(t)&&=\frac{e^{-(k+N\Gamma)t}}{2N(a^2+b^2)}
  [(a^2+b^2+\vert\Delta\vert^2)\cosh(bt)+ \\\nonumber
   &&(a^2+b^2-\vert\Delta\vert^2)\cos(at)-\\\nonumber
   &&2(b(\tilde{\omega}_0-\omega)-aA_-)\sin(at)+\\
  && 2(a(\tilde{\omega}_0-\omega)+bA_-)\sinh(bt)]
\end{eqnarray}
\begin{eqnarray}\label{rho2N+2}
 \nonumber \tilde{\rho}_{2,N+2}(t)&&=\frac{\varepsilon_{eff}e^{-(k+N\Gamma)t}}{N(a^2+b^2)}
  [\Omega(i\sin(at)+\sinh(bt))\\ &&+\Delta (\cosh(bt)-\cos(at))]
\end{eqnarray}
\begin{equation}\label{rhoN+2N+2}
  \tilde{\rho}_{N+2,N+2}(t)=\frac{2\varepsilon_{eff}^2e^{-(k+N\Gamma)t}}{N(a^2+b^2)}
  [\cosh(bt)-\cos(at)]
\end{equation}
with
\begin{eqnarray}\label{a}
 a&&=\{\frac{1}{2}\left[(\tilde{\omega}_0-\omega)^2+4\varepsilon_{eff}^2-A_-^2\right]\\\nonumber
&&
+\frac{1}{2}\left[((\tilde{\omega}_0-\omega)^2+4\varepsilon_{eff}^2-A_-)^2+
4(\tilde{\omega}_0-\omega)^2A_-^2\right]^{\frac{1}{2}}\}^{\frac{1}{2}},
\end{eqnarray}
\begin{eqnarray}\label{b}
 b&&=\{-\frac{1}{2}\left[(\tilde{\omega}_0-\omega)^2+4\varepsilon_{eff}^2-A_-^2\right]\\\nonumber
&&
+\frac{1}{2}\left[((\tilde{\omega}_0-\omega)^2+4\varepsilon_{eff}^2-A_-)^2+
4(\tilde{\omega}_0-\omega)^2A_-^2\right]^{\frac{1}{2}}\}^{\frac{1}{2}},
\end{eqnarray}
and $\Omega=a+ib=\frac{1}{2}[(\tilde{\omega}_0-\omega)^2-A_-^2+
4\varepsilon_{eff}^2+iA_-(\tilde{\omega}_0-\omega)]$,
$A_-=k-N\Gamma$, $\Delta=\tilde{\omega}_0-\omega+iA_-$ and
$\tilde{\omega}_0=\omega_0+\frac{N}{2}\Omega$. We wish to
emphasize that, on the basis of the block diagonal form exhibited
by $\tilde{\rho}_{AC}$, at a generic time instant $t$, the
transformed matter-radiation system is in a statistical mixture of
its vacuum density matrix and of an one-excitation appropriate
density matrix describing with certainty the storage of the
initial energy. Eqs.\ (\ref{rho22}) -\ (\ref{rhoN+2N+2}), giving
the explicit form of the time evolution of the combined physical
system, allow the exact evaluation of the mean value of any
physical observable of interest and, for instance, to follow the
entanglement formation or the progressive raising up of
decoherence effect in the matter-cavity subsystem.

\section{ Entanglement building up}
\label{sec:4} The circumstance that we succeed in finding the
explicit time dependence of the solution of the master equation\
(\ref{m-e-tr}), provides a lucky and intriguing occasion to
analyze in detail at least some aspects of how entanglement is
getting established in our exemplary enough multipartite system.
We wish indeed to point out that the question of how  to extend to
a generic $N$-partite physical system definition and measure of
entanglement built for bipartite systems, constitutes a topical
challenge involving many researchers
\cite{Wootters00,Wootters01,Wootters02,Wootters2002,Buzek03,Buzek2003,Partovi}.
Is is well understood from first principles that when many
subsystems of a multipartite system individually entangle a
prefixed one, the entanglement degree within each pair, anyhow
measured, is subjected to quantitative restrictions
\cite{Wootters00,Wootters2002,Buzek03,Buzek2003}. This, for
instance, implies that two maximally entangled subsystems of a
multipartite system are necessarily disentangled from any other
constituent units of the total system in an arbitrarily given pure
or not state. Of course, whatever the multipartite entanglement
definition is adopted, its occurrence is conceptually compatible
with a complete lack of partial entanglement of a given order for
example binary. On the contrary the existence of entanglement
between two specific subsystems has to be considered as a clear
symptom of entanglement in the multipartite system. Following this
line of reasoning we here therefore propose to study the time
evolution of the entanglement within all the possible binary
subsystems that is within each of the $\frac{N(N-1)}{2}$ pairs of
individual parts extractable from the $N$-partite set under
scrutiny. To this end we choose to evaluate the Wootter's
concurrence \cite{Wootters97,Wootters98} $C_{ij}(t)$ to
characterize quantitatively the formation of entanglement within
the $(i,j)-th$ pair of two-level atoms of our matter sample. Since
the dynamical problem exactly solved in the previous section is
invariant under the exchange of two arbitrary atoms, then one may
guess and indeed easily prove, that the reduced density matrix
\begin{equation}\label{ro2ij}
\rho^{(ij)}=Tr_{A_{ij}}\{U\tilde{\rho}_AU^{\dag}\}
\end{equation}
of the $(i,j)-th$ pair is {\sl structurally} independent from the
indices of two prefixed atoms meaning that the substitution of $i$
and $j$ with $i'$ and $j'$ respectively, exactly maps
$\rho^{(ij)}$ into $\rho^{(i'j')}$. The symbol $Tr_{A_{ij}}$ means
to trace over the atomic variables excluding the pair $(i,j)$
whereas $\tilde{\rho}_A=Tr_{cavity}\{\tilde{\rho}_{AC}\}$.

The concurrence $C_{ij}(t)$ is defined as
\begin{equation}
C_{ij}(t)=\max(0,\sqrt{\lambda_1^{(ij)}}-\sqrt{\lambda_2^{(ij)}}-\sqrt{\lambda_3^{(ij)}}
-\sqrt{\lambda_4^{(ij)}}),
\end{equation}
where $\lambda_q^{(ij)}$ $(q=1,...,4)$ are the decreasing-ordered
eigenvalues of the matrix
\begin{equation}
R^{(ij)}= \rho^{(ij)}\cdot \overline{\rho}^{(ij)}
\end{equation}
 where the spin
flipped matrix $\overline{\rho}^{(ij)}$ is given by
\begin{equation}\label{spinflip}
  \overline{\rho}^{(ij)}=\sigma^{(i)}_y \otimes \sigma^{(j)}_y
  (\rho^{(ij)})^\ast \sigma^{(i)}_y \otimes \sigma^{(j)}_y.
\end{equation}
 $(\rho^{(ij)})^\ast$ being the conjugate matrix of $\rho^{(ij)}$ \cite{Wootters97,Wootters98}.

In view of the invariance property of $\rho^{(ij)}$, it is not
difficult to persuade oneself that the eigenvalues of
$R^{(ij)}(t)$ are pair-independent, which, as a consequence,
implies $C_{ij}(t)=C_{N-1,N}(t)$ for any $i,j=1,2,...,N$ $(i<j)$.
Let's then start by extracting the expression of
$\rho^{(N-1,N)}(t)$. Tracing in accordance with eq.\ (\ref{ro2ij})
yields
\begin{eqnarray}\label{rorid1}
  &&\rho_{N-1,N}(t)=a|-\rangle_{N-1}|-\rangle_N\;_{N-1}\langle -|_{N}\langle-|\\\nonumber &+&
  b|+\rangle_{N-1}|-\rangle_N\;_{N-1}\langle +|_{N}\langle-|+
  c|-\rangle_{N-1}|+\rangle_N\;_{N-1}\langle -|_{N}\langle+|\\\nonumber &+&
  d|+\rangle_{N-1}|-\rangle_N\;_{N-1}\langle -|_{N}\langle+|+
  e|-\rangle_{N-1}|+\rangle_N\;_{N-1}\langle +|_{N}\langle-|,
\end{eqnarray}
where
\begin{eqnarray}\label{coef}
  && \nonumber
  a=(1-\frac{2}{N})\tilde{\rho}_{22}(t)+(N-3+\frac{2}{N})\tilde{\rho}_{N,N}(t)+
  \\&&+\tilde{\rho}_{N+2,N+2}(t)\\
  &&b=c=\frac{N-2}{N}\tilde{\rho}_{22}(t)+\frac{N-1}{N}\tilde{\rho}_{N,N}(t)\\
  &&d=e=\frac{1}{N}(\tilde{\rho}_{22}(t)-\tilde{\rho}_{N,N}(t))
\end{eqnarray}
It is now easy to construct and diagonalize $R^{(N-1,N)}(t)$
finally getting for any pair $(i,j)$ with $i<j$
\begin{eqnarray}\label{conc}
 C_{ij}(t)=\frac{2|\tilde{\rho}_{22}(t)-\tilde{\rho}_{N+1,N+1}(t)|}
 {N\sum_{i=2}^{N+2}\tilde{\rho}_{ii}(t)}.
\end{eqnarray}
The $c-$function $C_{ij}(t)$ is displayed in figures (1)-(3) in
correspondence to $N=2,10$ and 100 respectively using reasonable
values for the involved parameters \cite{Pellizzari}. It
represents the conditional concurrence characterizing the temporal
evolution of bipartite entanglement under the hypothesis that the
only photon, initially injected in the system has not escaped
because of loss mechanisms. The fact that $C_{ij}(t)$ is different
from zero at any time instant $t>0$ whatever the pair is,
undoubtedly reflects the existence of a process giving rise to the
entanglement formation inside the $N-$partite system. On the other
hand since each atom turns out to be entangled with all the
others, the concurrence of each couple of atoms is monotonically
decreasing with $N$. As already mentioned at the beginning of this
section, this behaviour reflects nothing but the expected
reduction of atom-atom entanglement due to the increase in the
number of possible entangled couples. Albeit $C_{ij}(t)$ tends to
vanish when the number $N$ of atoms goes to infinity, we find the
remarkable result that the {\sl total binary concurrence }
\begin{eqnarray}\label{conc}
 \nonumber&&C_{BT}(t)\equiv\sum_{i>j}C_{ij}(t)=\\&&
 =\frac{(N-1)|\tilde{\rho}_{22}(t)-
 \tilde{\rho}_{N+1,N+1}(t)|}{\sum_{i=2}^{N+2}\tilde{\rho}_{ii}(t)},
\end{eqnarray}
exhibits an oscillatory time behaviour with a decreasing amplitude
around a time-dependent mean value monotonically tending toward
the stationary value 1, whatever $N$ is (see figures (4)-(6)).
This dynamical property is of relevance because examples of
multipartite system states manifestly entangled, for which
$C_{BT}=0$ may be provided \cite{Buzek03}. Stated another way,
eq.\ (\ref{conc}) tell us that the not vanishing contribution of
the total binary entanglement to the formation of entanglement
within our multipartite dynamical problem does not scale with $N$,
reflecting that the $N-$decrease of each $C_{ij}(t)$ is well
compensated by the quadratic $N-$increase of the number of pairs.
We thus claim that the behaviour of $C_{BT}(t)$, and in particular
its asymptotic tendency, provides, in our situation, a peculiar
feature helping to describe and understand some aspects of the
entanglement formation in our multipartite system.
\begin{figure}
 \vspace{0.1 cm}
 \includegraphics{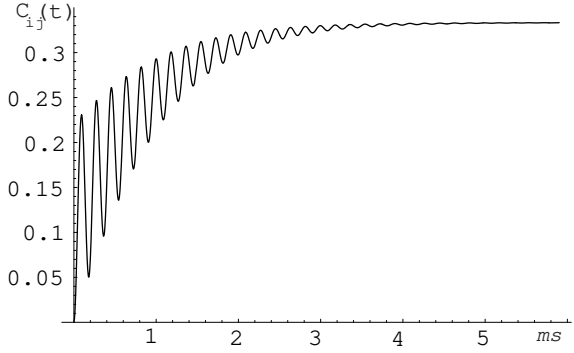}
  \caption{Conditional binary concurrence $C_{ij}(t)$ in correspondence to $N=3$,
  $\varepsilon=10^{5}Hz$,
  $k=10^{4}Hz$, $\Gamma=10^{3}Hz$ and
  $\omega=\tilde{\omega}_0=10^{14}Hz$ }
  \label{Conc3}
\end{figure}

\begin{figure}
 \vspace{0.1 cm}
 \includegraphics{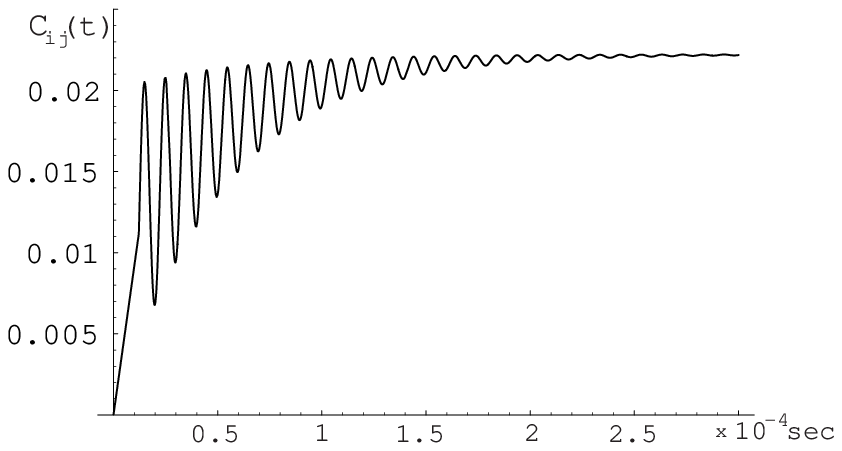}
  \caption{Conditional binary concurrence $C_{ij}(t)$ in correspondence to $N=10$,
  $\varepsilon=10^{5}Hz$,
  $k=10^{4}Hz$,$\Gamma=10^{3}Hz$ and
  $\omega=\tilde{\omega_0}=10^{14}Hz$ }
  \label{Conc10}
\end{figure}

\begin{figure}
 \vspace{0.1 cm}
 \includegraphics{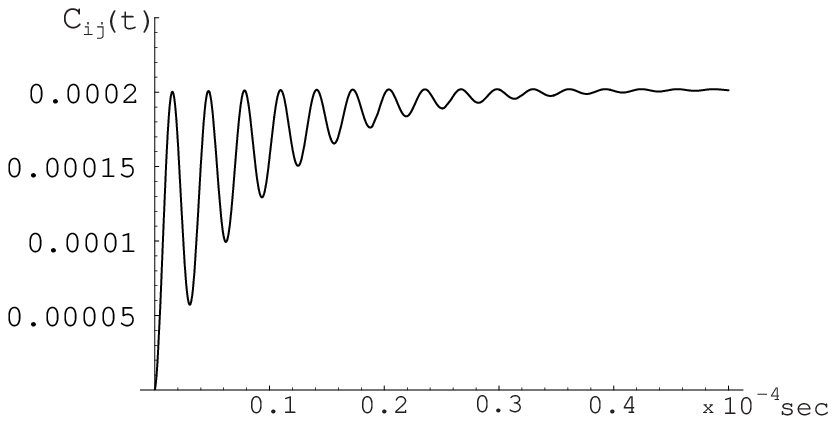}
  \caption{Conditional binary concurrence $C_{ij}(t)$ in correspondence to $N=10$,
  $\varepsilon=10^{5}Hz$,
  $k=10^{4}Hz$,$\Gamma=10^{3}Hz$ and
  $\omega=\tilde{\omega_0}=10^{14}Hz$ }
  \label{Conc100}
\end{figure}

\begin{figure}
 \vspace{0.1 cm}
 \includegraphics{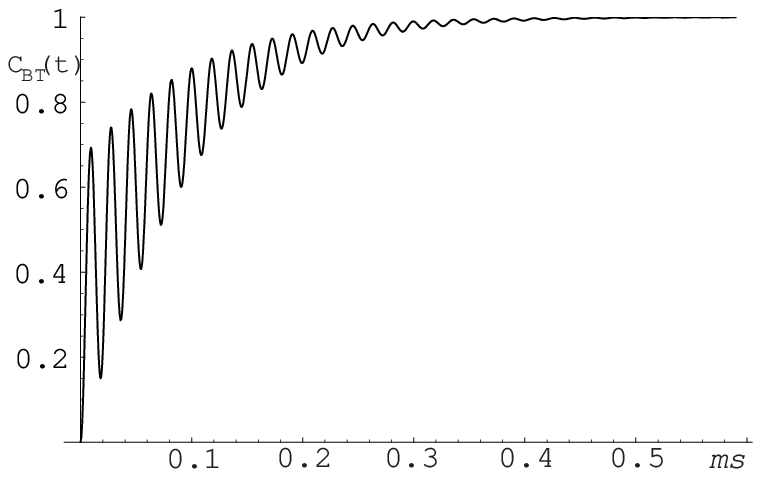}
  \caption{Total binary concurrence $C_{BT}(t)$ in correspondence to $N=3$,
  $\varepsilon=10^{5}Hz$,
  $k=10^{4}Hz$,$\Gamma=10^{3}Hz$ and
  $\omega=\tilde{\omega_0}=10^{14}Hz$}
  \label{Concsomma3}
\end{figure}

\begin{figure}
 \vspace{0.1 cm}
 \includegraphics{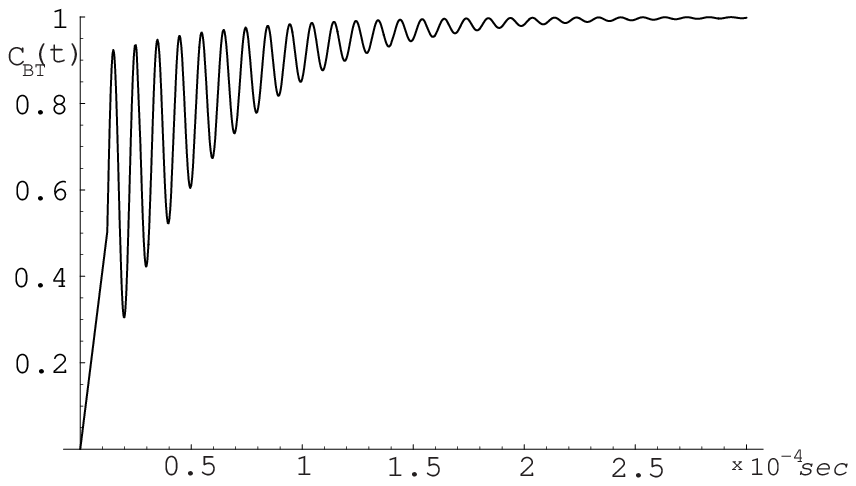}
  \caption{Total binary concurrence $C_{BT}(t)$ in correspondence to $N=10$,
  $\varepsilon=10^{5}Hz$,
  $k=10^{4}Hz$,$\Gamma=10^{3}Hz$ and
  $\omega=\tilde{\omega_0}=10^{14}Hz$}
  \label{Concsomma10}
\end{figure}

\begin{figure}
 \vspace{0.1 cm}
 \includegraphics{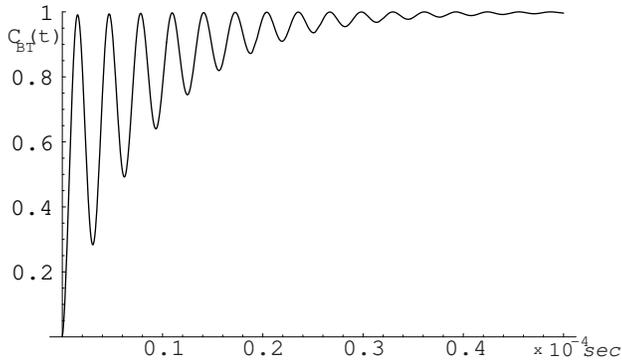}
  \caption{Total binary concurrence $C_{BT}(t)$ in correspondence to $N=100$,
  $\varepsilon=10^{5}Hz$,
  $k=10^{4}Hz$,$\Gamma=10^{3}Hz$ and
  $\omega=\tilde{\omega_0}=10^{14}Hz$}
  \label{Concsomma100}
\end{figure}

\section{The asymptotic form of $\tilde{\rho}_{AC}$ }\label{sec:2}
It is of particular relevance that for

\begin{equation}\label{tauac}
t\gg (k+N\Gamma)^{-1}\equiv\tau_{AC}
\end{equation}
the correspondent asymptotic form assumed by $\tilde{\rho}_{AC}$
is time independent and such that the probability of finding
energy in the effective JC subsystem exactly vanishes. Taking into
account the easily demonstrable inequality $b<\tau_{AC}^{-1}$, it
is immediate to convince oneself that for $t\gg \tau_{AC}$, the
eq.\ (\ref{matrix1}) assumes the following diagonal asymptotic
form
\begin{eqnarray}\label{matrix2}
\tilde{\rho}_{AC}(t\gg \tau_{AC})= \left( \begin{array}{c c c c c
c}

 \frac{1}{N} & 0 & 0 & ... & 0 & 0 \\

0 & 0 & 0 & ... & 0 & 0 \\

0 & 0 & \frac{1}{N} & ... & 0 & 0 \\

\vdots & \vdots & \vdots & \vdots\vdots\vdots & \vdots & \vdots \\

0 & 0 & . & ... & \frac{1}{N} & 0 \\

0 & 0 & . & ... & 0 & 0
\end{array}
\right)
\end{eqnarray}

Starting from eq.\ (\ref{matrix2}) and coming back to the old
representation, it is possible to give at any time instant $t$ the
exact solution $\rho_{AC}$ for the reduced density matrix of the
system under scrutiny. Taking into account that the unitary
operator $U$ is time independent, in view of eq.\ (\ref{matrix2})
\begin{eqnarray}
\rho_{AC}=U\tilde{\rho}_{AC}(t\gg\tau_{AC})U^{\dag}
\end{eqnarray}
is time independent too.  In fact we find that for $t\gg
\tau_{AC}$, the reduced density matrix can be written in the
compact form
 \begin{eqnarray}\label{roAC}
  \rho_{AC}=\frac{1}{N}\vert \beta_1\rangle\langle \beta_1
  \vert+\frac{N-1}{N^2}\sum_{h=2}^{N+1} \vert \psi_T^{(h)}\rangle\langle \psi_T^{(h)}
  \vert \\\nonumber\equiv\frac{1}{N}\vert \beta_1\rangle\langle \beta_1
  \vert+\frac{N-1}{N}\rho_{mix}
\end{eqnarray}
where
\begin{eqnarray}\label{psiT}
 \nonumber \vert \psi_T^{(h)} \rangle &&\equiv \frac{1}{\sqrt{N(N-1)}} \{(N-1)
  \vert\beta_h \rangle \\&& -\sum_{j\neq h,j=2}^{N+1}\vert \beta_j
  \rangle \}\equiv\vert 0\rangle\vert\varphi_T^h\rangle.
\end{eqnarray}

It is worth noting that each normalized state
$\vert\varphi_T^{(h)}\rangle$ given by eq.\ (\ref{psiT}), defines
a particular subradiant Dicke state. It is indeed possible to
prove that, whatever $N$ and $h$ are, the states $\vert
\varphi_T^{(h)} \rangle$ are common eigenstates of $S_z$ and $S_-$
pertaining to eingevalues $-\frac{N-2}{2}$ and $0$ respectively.
This property is a sufficient condition to claim that $\vert
\psi_T^{(h)} \rangle$ is eigenstate of $\mathbf{S}^2$ too, having
the form $\vert S, -S \rangle $ with $S=\frac{N-2}{2}$.  This
means that
\begin{equation}\label{s2romix}
 \mathbf{S^2}\rho_{mix}\mathbf{S^2}=(S(S+1))^2\rho_{mix}
\end{equation}
\begin{equation}\label{szromix}
 S_z\rho_{mix}S_z=S^2\rho_{mix}
\end{equation}
and that each $\vert \psi_T^{(h)} \rangle$ defines an example of
subradiand or trapped state \cite{Benivegna89,Dicke,Dorris,Tavis}.
Thus the result expressed by eq.\ (\ref{roAC}) suggests that a
statistical mixture of \textit{stationary subradiant} Dicke states
of the atomic sample, having  well defined values of
$\mathbf{S^2}$ and $S_z$, can be generated, at least in principle,
putting outside the cavity single photon detectors allowing us to
continuously monitor the decay of the system through the two
possible channels (atomic and cavity dissipation)
\cite{Carmichael}. Eq.\ (\ref{roAC}), indeed, clearly shows that,
reading out the detectors state at $\bar{t}\gg \tau_{AC}$, if no
photons has been emitted, then, as a consequence of this
measurement outcome, our system is projected into the state
$\frac{1}{N}\sum_{h=2}^{N+2}\vert \psi_T^{(h)} \rangle\langle
\psi_T^{(h)} \vert$.

Stated another way, successful measurements, performed at large
enough time instants $t$, generates an uncorrelated state of the
two atomic and cavity field subsystems, leaving the matter sample
in the statistical mixture of Dicke states $\vert \varphi_T
^{(h)}\rangle$, satisfying eqs.\ (\ref{s2romix}) and\
(\ref{szromix}). On the basis of the analysis reported in the
previous section it is possible to state that such a statistical
mixture is entangled.

Let's  finally observe that the probability $P$ that at
$t=\bar{t}$ the excitation is still contained in the atomic system
increases with the number $N$ of atoms, being $P=1-\frac{1}{N}$,
as immediately deducible from eq.\ (\ref{roAC}).

\section{Conclusive Remarks}
\label{sec:3} Summing up, in this paper we have exactly solved the
dynamics of $N$ identical atoms resonantly interacting with a
single mode cavity, taking into account from the very beginning
the presence of both the resonator losses and the atomic
spontaneous emission. We have moreover supposed that only one
excitation is initially injected into the system of interest and
that the atoms are located in such a way to experience the same
cavity field.

From a mathematical point of view \textit{the novelty of our
results} is the presentation of an exact way to solve the master
equation eq.\ (\ref{m-e}) of the system, based on the unitary
transformation accomplished by the operator U given by eq.\
(\ref{U}).  In the new representation associated to such a
specific $U$, the differential equations governing the temporal
behavior of the density matrix elements $\tilde{\rho}_{hj}$,
become indeed much simpler to solve if compared with the ones
ruling the temporal behavior of $\rho_{hj}$.  This circumstance
stems from the fact that in the new representation only one atom
is at the same time coupled both to the cavity field and to the
quantized electromagnetic modes of the thermal bath. The
decoupling of the $N-1$ atoms is a direct consequence of the
permutational symmetry properties acquired by the matter subsystem
under the assumed point-like model condition.

From a physical point of view \textit{the new results reported in
this paper} have the merit of providing the key for transparently
interpreting the origin of the asymptotic occurrence of collective
subradiant Dicke behavior of the matter subsystem. The analysis
and the discussion presented in section 4, highlighting some
features of the entanglement formation process, legitimate the
claim that the asymptotic condition toward which our physical
system is guided by the loss mechanisms exhibits entanglement. It
is of relevance to notice that the form of the stationary
conditional state $\rho_{mix}$ appearing in eq.\ (\ref{roAC}) is
independent from both $\Gamma$ and $k$, while the exponential
tendency toward the stationary condition is governed by the rate
$k+N\Gamma$. This means that just the presence of only one loss
channel is sufficient to address the same asymptotic radiation
trapping condition, \textit{even if the transient duration is
characterized by a different time decay constant}. We wish in
addition to emphasize that if the microscopic model neglects
spontaneous atomic decay, cooperative effects occur even if the
atomic sample is spatially dispersed
\cite{Kozierowski,Shore,Kudryavstev,Phoenix}. On the contrary, the
closeness of the $N$ atoms is a necessary requirement when the
complete Hamiltonian model\ (\ref{Htot}) is used (regardless of
how bad the cavity is) in order that a robust entanglement may be
conditionally reached in the matter subsystem. If indeed the
distance among the $N$ atoms is larger than the radiation
wavelength, collective behavior stemming from the interaction of
each atom with a common bath, disappear with the consequence that
the probability of finding in the system the initial energy goes
toward zero with time. We thus may state that the key for trapping
the energy in the atomic sample, inducing a stationary collective
Dicke behavior, is the closeness among the atoms. We wish to
conclude presenting some remarks concerning the experimental
relevance of the problem discussed in this paper. We begin by
observing that, in view of eq.\ (\ref{tauac}), the value of
$\tau_{AC}$ correspondent to $k=10^4Hz$ and $\Gamma=10^3Hz$
\cite{Pellizzari} becomes much less than $10^{-2}$ whatever $N$
is. The experimental implementation of the specific conditions
envisaged in our paper thus require the ability of locating for a
time of the order of $10^{-2} sec$, $N$ atoms in an enough small
region within an optical resonator. In particular the distance
between two arbitrarily chosen atoms of our matter sample has to
be much less than  $500 nm$. The more and more growing
technological successes registered in the last few years in the
confinement of individual atoms \cite{Balykin,Chu,1atom,1atom2} or
clouds of identical atoms \cite{Balykin,Chu,Ozeri} with high
spatial resolution, suggest that implementing our conditions in
the near future is in the grasp of the experimentalists.To enforce
our claim it is appropriate and relevant to quote the paper of
Ozeri et al \cite{Ozeri} wherein the authors experimentally
demonstrate the possibility of confining in a blue-detuned optical
trap a sample of $10^5$ Rubidium atoms in a region of $0.1\div 30
nm$ for $0.3 sec$ at a temperature of $24 \mu K$.

\end{document}